\documentclass[twocolumn]{aa}
\usepackage{graphicx}
\usepackage{txfonts}
\begin{document}

\title{ The Radial Distribution of the Two Stellar Populations in NGC~1851
	\thanks{Based on observations with the NASA/ESA {\it Hubble
Space Telescope}, obtained at the Space Telescope Science Institute,
which is operated by AURA, Inc., under NASA contract NAS 5-26555,
under GO-11233.}}


\author{
A. \,P. \,Milone\inst{1},
P. \,B. \,Stetson\inst{2},
G. \,Piotto\inst{1},
L. \,R. \,Bedin\inst{3},
J. \,Anderson\inst{3},
S. \,Cassisi\inst{4}, and
M. \,Salaris\inst{5}
}

\offprints{A. \ P. \ Milone}

\institute{
Dipartimento  di   Astronomia,  Universit\`a  di
  Padova, Vicolo dell'Osservatorio 3, Padova, I-35122, Italy
\and
Dominion Astrophysical Observatory, Herzberg Institute of Astrophysics,
National Research Council, 5071 West Saanich Road, Victoria, BC V9E 2E7, Canada
\and
Space Telescope Science Institute, 3700 San Martin Drive,
Baltimore, MD 21218, USA
\and
INAF-Osservatorio Astronomico di Collurania,
Via M. Maggini, Teramo I-64100, Italy
\and
 Astrophysics Research Institute, Liverpool John Moores University,
 Twelve Quays House, Egerton Wharf, Birkenhead CH41 1LD, UK
%
           }

\date{Received Xxxxx xx, xxxx; accepted Xxxx xx, xxxx}
%

\abstract{
We have analyzed ACS/WFC and WFPC2 images from {\it HST}, as well as
ground-based data to study the radial distribution of the double sub-giant
branch (SGB) recently discovered in the Galactic globular cluster NGC 1851.
We found that the SGB split can be followed all the way from the cluster center
out to at least 8\arcmin\ from the center.  Beyond this distance out to the
tidal radius at $\sim 11.7$ arcmin, there are simply too few SGB stars to
identify the sequences.
The number ratio of the bright SGB to the faint SGB stars shows no
significant radial trend.
Furthermore, we have found that the ratio of blue horizontal-branch (HB)
      stars plus RR Lyrae to the red HB stars also remains
      constant from the cluster center to the outer envelope.
}

\titlerunning{NGC~1851 SGB's gradient}
\authorrunning{Milone et al.}

\maketitle

%
\section{Introduction}
\label{introduction}
%

Precise  Hubble  Space  Telescope   ({\it HST})  photometry
has
provided
evidence that  NGC~1851 hosts two  distinct sub-populations of stars
(Milone et
al.\ 2008) as  indicated by  a clear bifurcation of  the
sub-giant branch (SGB)
in  its  color  magnitude diagram  (CMD).    This
discovery has  sparked new interest    in this object
and, consequently,
numerous efforts aimed at  better  understanding how this cluster  formed
and evolved.

Milone et  al.\ (2008) suggest  that  two
star-formation  episodes
delayed by  about one Gyr could explain  the observed split of  the SGB.
As an  alternative scenario, Cassisi et al.\  (2008)  suggest that the
SGB split  can  be   explained
by the  presence  of   two  stellar
populations,  one   with
normal $\alpha$-element enhancement,
and the other characterized
by a peculiar CNONa chemical pattern with C+N+O
abundance increased by a factor of $\sim$ 2.
Two such populations
 could account  for the observed bright
and   faint  SGB (hereafter   bSGB   and fSGB), respectively,  without
requiring any significant age difference.

Interestingly, this      latter scenario seems      to be supported by
early spectroscopic measurements (Hesser et  al.\  1982)
which indicate  the
presence of two groups:  CN-strong and CN-weak stars.
In addition,  the work of  Calamida et al.\  (2007) shows that  in the
Str\"omgren  ($m_{1}$,  $u-y$) CMD,  the  red  giant  branch (RGB)  of
NGC~1851
splits into two sequences that, like the SGB split,
       can be explained by two populations with different CN
       abundances.

A more recent  spectroscopic investigation by  Yong \& Grundahl (2008)
determined   the  chemical  composition  for  eight   bright giants in
NGC~1851.   Their    analysis revealed  large   star-to-star
light-element-abundance
differences of the elements Zr and La.  These $s$-process elements are
correlated with Al and anticorrelated with O.  Furthermore, the Zr and
La abundances  appear  to peak around  two  distinct values.  Yong  et
al.  (2009) show  that  the C+N+O abundance   exhibits  a large spread
($\sim$ 0.6 dex), giving further support to the  Cassisi et al. (2008)
scenario. They also found  a correlation of  Na, Al, Zr, La  abundance
with C+N+O, as expected in the scenario in which intermediate mass AGB
stars  are  responsible for   globular-cluster light-element abundance
variation.

Another important peculiarity of NGC~1851 is that it is one of only a
few examples of a bimodal horizontal branch (HB) cluster.
On the basis of detailed  numerical simulations, Salaris et al. (2008)
found that  it is  possible to account  simultaneously for  the various
empirical constraints such as the HB morphology, the star counts along
the HB  as well as  the ratio between  faint SGB stars and  bright SGB
ones,if all the  upper SGB stars  and a small   fraction of lower SGB
stars evolves into  the  red HB, while most  of  the lower   SGB stars
populates the blue HB (including the RR Lyrae variables).

Therefore   it is  tempting   to associate  the  bSGB   stars with the
CN-normal, $s$-process-normal  stars and  with the  red  HB, while the
fainter     SGB     should     be     populated      by     CN-strong,
$s$-process-element-enhanced stars  which  should  evolve  mainly
onto the blue HB.  This scenario implies that fSGB stars correspond to
the second generation  and that they   formed from material  processed
through a first generation of stars.

The study of the spatial distribution of the two stellar populations
associated with the double SGB occupies a pivotal role in the current
research on multiple stellar populations in GCs.
Milone et al.\ (2008) studied a CMD containing stars with radial
distances smaller than $\sim$2.5 arcmin from the cluster center.
  In this paper, we extend the analysis to the
outermost regions of the cluster. Our
radial extension covers the whole cluster from the center
to beyond the tidal radius ($r_{tidal}$=11.7 arcmin, Harris \ 1996, 2003).

A first attempt at constraining the radial distribution of the two
stellar populations associated with the double SGB comes from the
recent work by Zoccali et al.\ (2009). They analyzed VLT-FORS $V$, $I$
images, of the South West quadrant of the cluster with the aim of
following the extent of the double SGB from $\sim$1 to $\sim$13
arcmin from the cluster center.
Zoccali  et al. \ (2009)  claimed that  the  percentage of fSGB stars,
which  is $\sim$45\% in  the  innermost region decreases at  $\sim$1.5
arcmin from the center and completely disappears at $\sim$ 2.4 arcmin.
The more extensive and higher-precision data set presented
             below does not confirm the results of Zoccali et al.\ (2009).

The plan of the paper is  as follows.  In the next section we describe
the data  and the data reduction.   The CMDs from  different data sets
are presented in Sect.~\ref{CMD}.   In Sect.~\ref{RaD} we describe the
criteria  used  to define  the  NGC~1851  sub-populations  and how  we
measured  their relative  frequency at  several radial  distances.  We
present  the radial  distribution of  the two  stellar  populations in
NGC~1851 and discuss our results in Sect.~\ref{discussion}.

   \begin{figure}[ht!]
   \centering
   \includegraphics[width=8.25 cm]{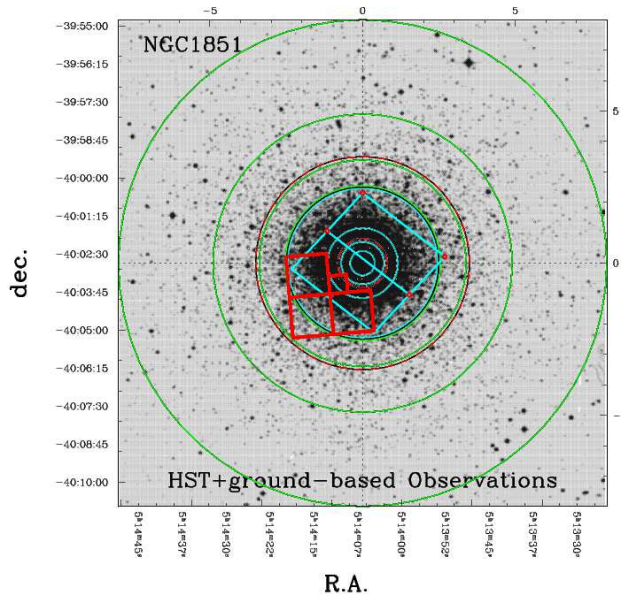}
      \caption{
For the innermost 8 arcmin, we show the
footprints of the {\it HST}-images used in this paper
superposed on a DSS image.
               In cyan ACS/WFC (circles highlight the corners of the
               first chip of ACS), in red WFPC2.
               Large
               circles show the different radial bins described
               in Section \ref{RaD}. The different colors
               refer to different data-sets: cyan for ACS/WFC,
               red for WFPC2, and green for
               ground-based data-set.
	}
         \label{fields}
   \end{figure}
%

\begin{table}[ht!]
\caption{Description of the
{\it HST}
data sets used in this paper. }

\scriptsize {
\begin{tabular}{lcccl}
\hline
\hline
 INSTR &  DATE & N$\times$EXPTIME & FILT  & PROGRAM (PI) \\
\hline
 & & & &  \\
 ACS/WFC   & May 01 2006   &  20s$+$5$\times$350s           & F606W  &  10775 ( Sarajedini) \\
 ACS/WFC   & May 01 2006   &  20s$+$5$\times$350s           & F814W  &  10775 ( Sarajedini) \\
 & & & &  \\
 WFPC2     & Nov 07 2007   & 10$\times$230s                 & F450W  &  11233 ( Piotto) \\
 WFPC2     & Nov 07 2007   & 5$\times$200s$+$5$\times$230s  & F814W  &  11233 ( Piotto) \\
 & & & &  \\
\hline
\hline
\end{tabular}
}
\label{tabdata}
\end{table}

%
\section{Observation and data reduction}
\label{data}
%

In order to study the radial distribution of SGB and HB stars in NGC~1851, we
considered three distinct data sets.

To probe the most crowded regions of the cluster we took advantage of
the high resolving power of {\it HST}, using images collected with the
Wide Field Channel (WFC) of the Advanced Camera for Survey (ACS)
and with the Wide Field Planetary Camera 2 (WFPC2).
NGC~1851 is not in a particularly dense region of the Galaxy ($l^{\hbox{II}} = 245^\circ,
b^{\hbox{II}} = -35^\circ$),
and ground-based observations can provide photometry outside the
crowded
central region with precision comparable to that of the more central regions
observed with the {\it HST} cameras.

A brief description of the
{\it HST}
images used in this work is given in
Table~\ref{tabdata}, while Fig.~\ref{fields} shows a finding-chart of
their footprints.

\subsection{The ACS/WFC data-set}
The {\it HST} ACS/WFC images come from  GO-10775 (PI:\ Sarajedini, see also Sarajedini et al. \ 2007) and
were presented   in Milone  et  al. \  (2008); we have used
 the output of the reduction described in
               Anderson et al. (2008).  In brief, the procedure
               analyzes
all  the exposures of each cluster
simultaneously  to generate a   single     list of stars  for     each
field. Stars  are measured  independently in each  image  by using the
best available PSF models from Anderson \& King (2006).

This routine was designed  to work well  in both crowded and uncrowded
fields, and it  is  able to  detect   almost every  star that  can  be
perceived by  eye. It takes advantage of  the many independent dithered
pointings  of  each  scene and   the knowledge  of   the PSF  to avoid
including artifacts  in the list.  Calibration of ACS  photometry into
the Vega-mag system  was performed following  recipes in Bedin et al.\
(2005) and using the zero points given in Sirianni et al.\ (2005).

\subsection{The WFPC2 data-set}
The {\it HST} WFPC2 images come from GO-11233 (PI:\ Piotto), a
proposal specifically dedicated to photometric detection of multiple
populations.

The WFPC2 images have been reduced following the method of Anderson
\& King (2000), which is based on effective-point-spread-function
fitting.  We corrected for the $34^{th}$ row error in  WFPC2
CCDs (see Anderson \& King \ 1999 for details) and used the best
distortion solution
available, as
given by Anderson \& King (2003).
Photometric calibration has been done according to the Holtzman et
al.\ (1995) Vega-mag flight system for WFPC2 camera.

\subsection{The  ground-based data-set}

The ground-based  data are taken  from the image archive maintained by
one of  us (Stetson \ 2000).   The observations used here  include 545
images from 14 observing runs with the Max  Planck 2.2m telescope, the
CTIO 4m, 1.5m, and 0.9m telescopes, and the Dutch 0.9m telescope on La
Silla.  Any given star may have as many as 69 independent measurements
in the $B$ filter, 78 in $V$, and 56 in $I$; among  these, 62, 70, and
56,  respectively, were taken on occasions  that  were judged to be of
photometric quality.  These data were  reduced following the  protocol
outlined in  some  detail  in Stetson   \  (2005).  We   have complete
photometric   coverage of the    cluster field  out to   a  radius  of
14.5~arcmin, and partial coverage to 25.9~arcmin, but we will restrict
our discussion here to the  area within 12.5~arcmin, which is slightly
larger than the tidal radius of 11.7~arcmin.

%
\section{The CMDs}
\label{CMD}
%

\begin{figure*}[ht!]
   \centering
   \includegraphics[width=8.25 cm]{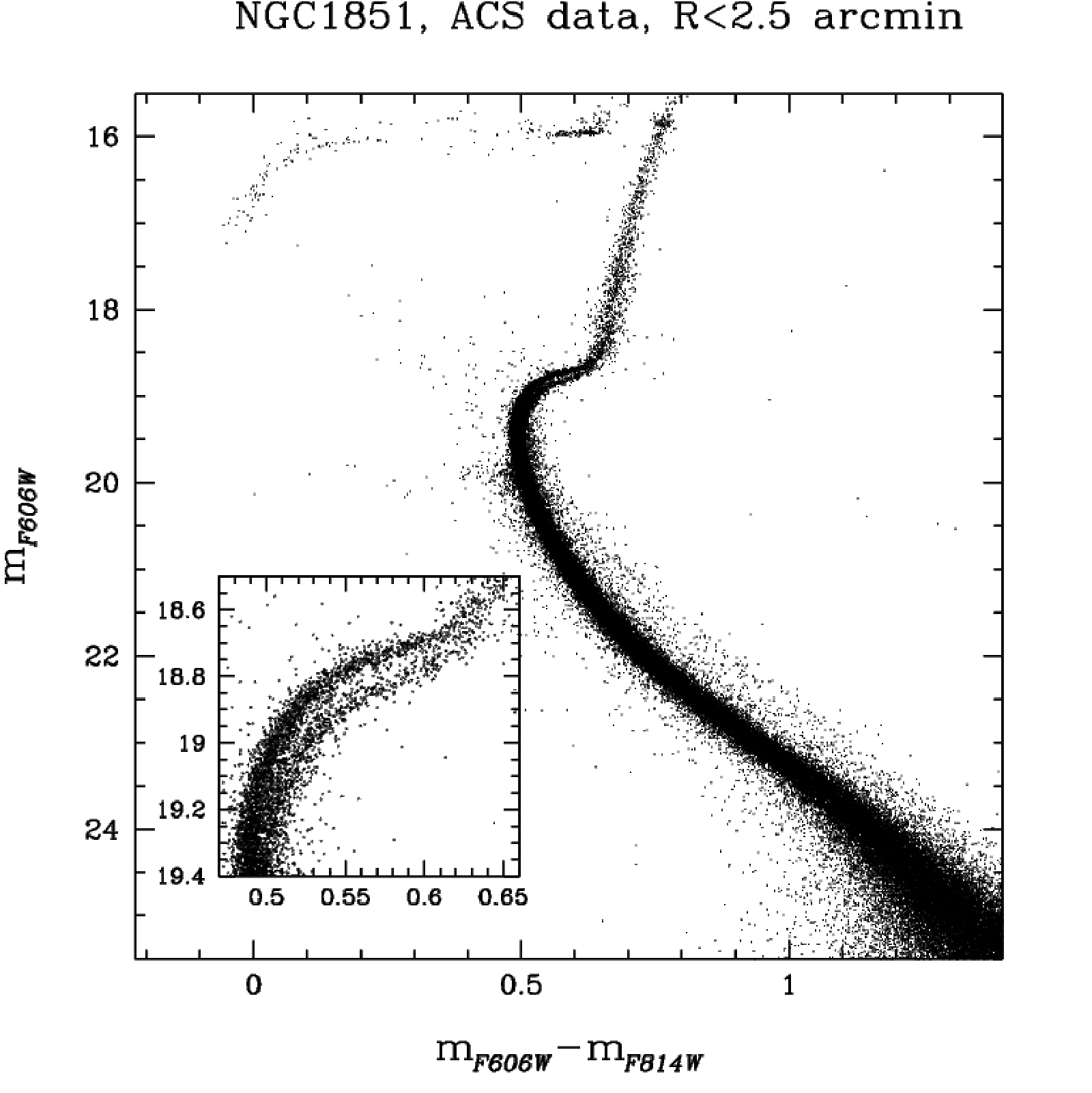}
   \includegraphics[width=8.25 cm]{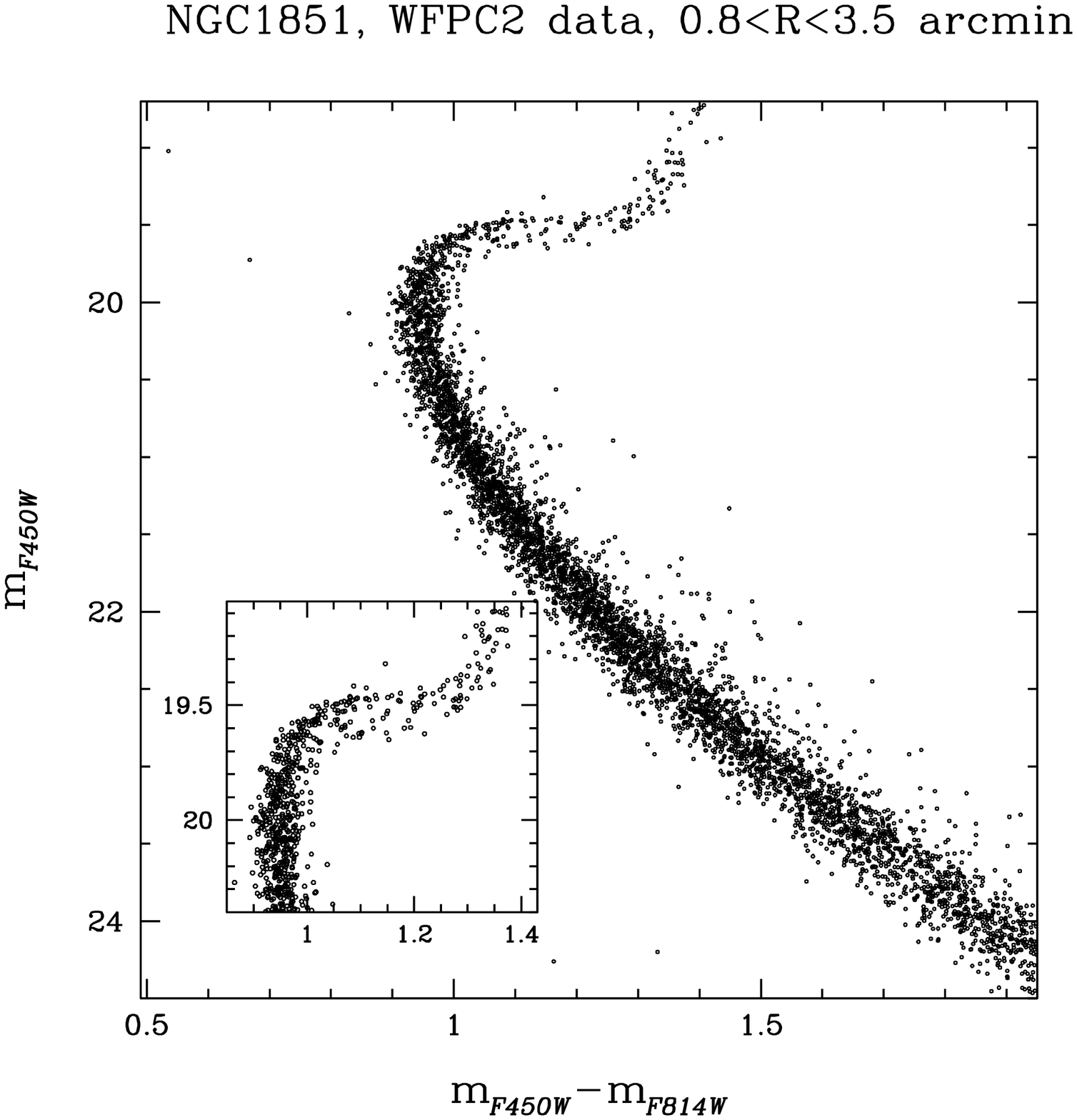}
      \caption{ CMD of  NGC~1851 from ACS ($left$) and WFPC2 ($right$) data.
	The inset show  a zoom around   the SGB  region.  }
         \label{SB}
   \end{figure*}

Figure~\ref{SB} shows the CMDs  from {\it HST} observations that cover
the densest  regions  of the cluster.  The  left  panel shows the  CMD
already published by Milone et al.\ (2008), and the right panel shows
the newly derived CMD from the WFPC2 data described in Sect.\ 2.2.

\begin{figure*}[ht!]
   \centering
   \includegraphics[width=8.25 cm]{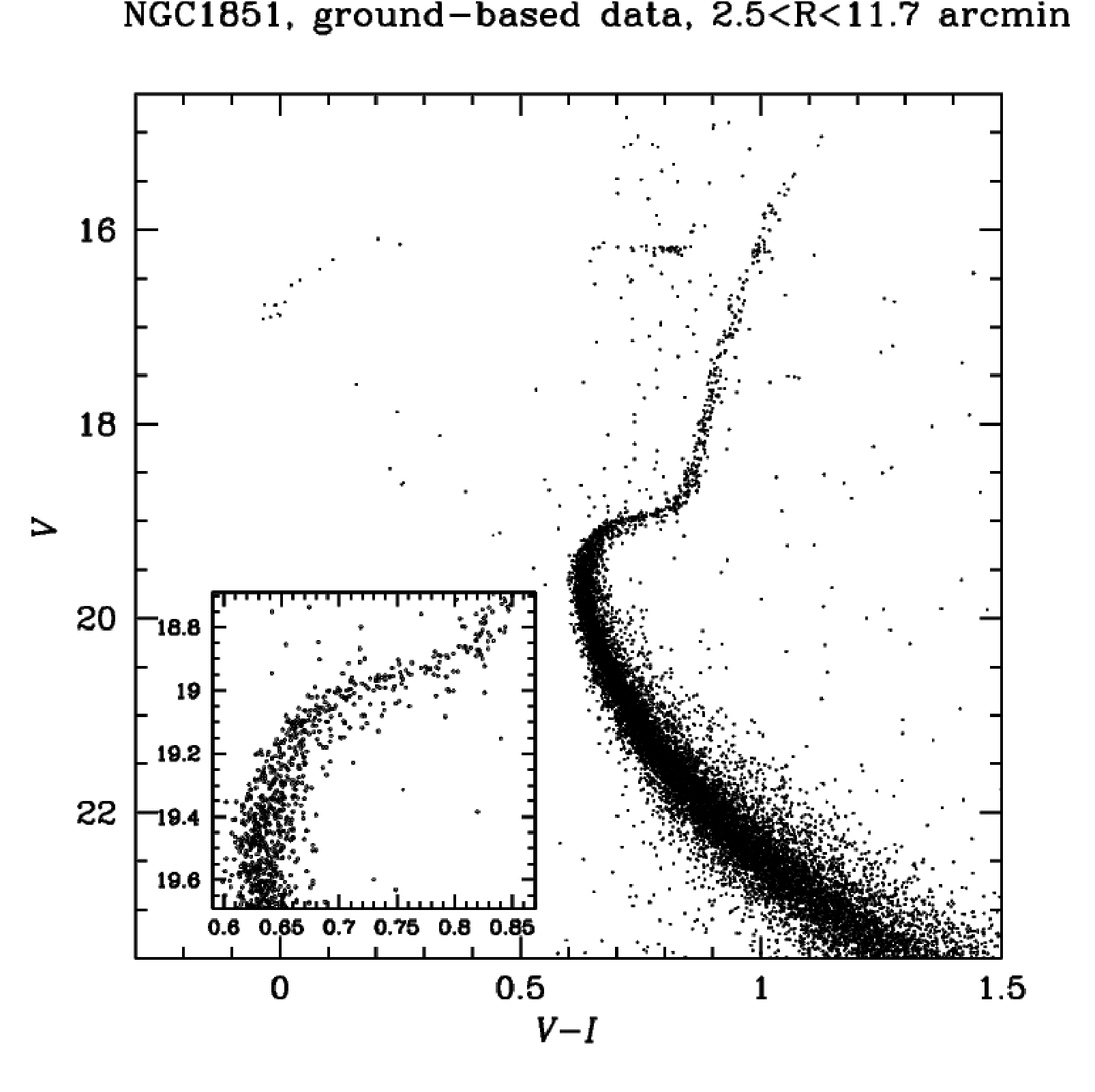}
   \includegraphics[width=8.25 cm]{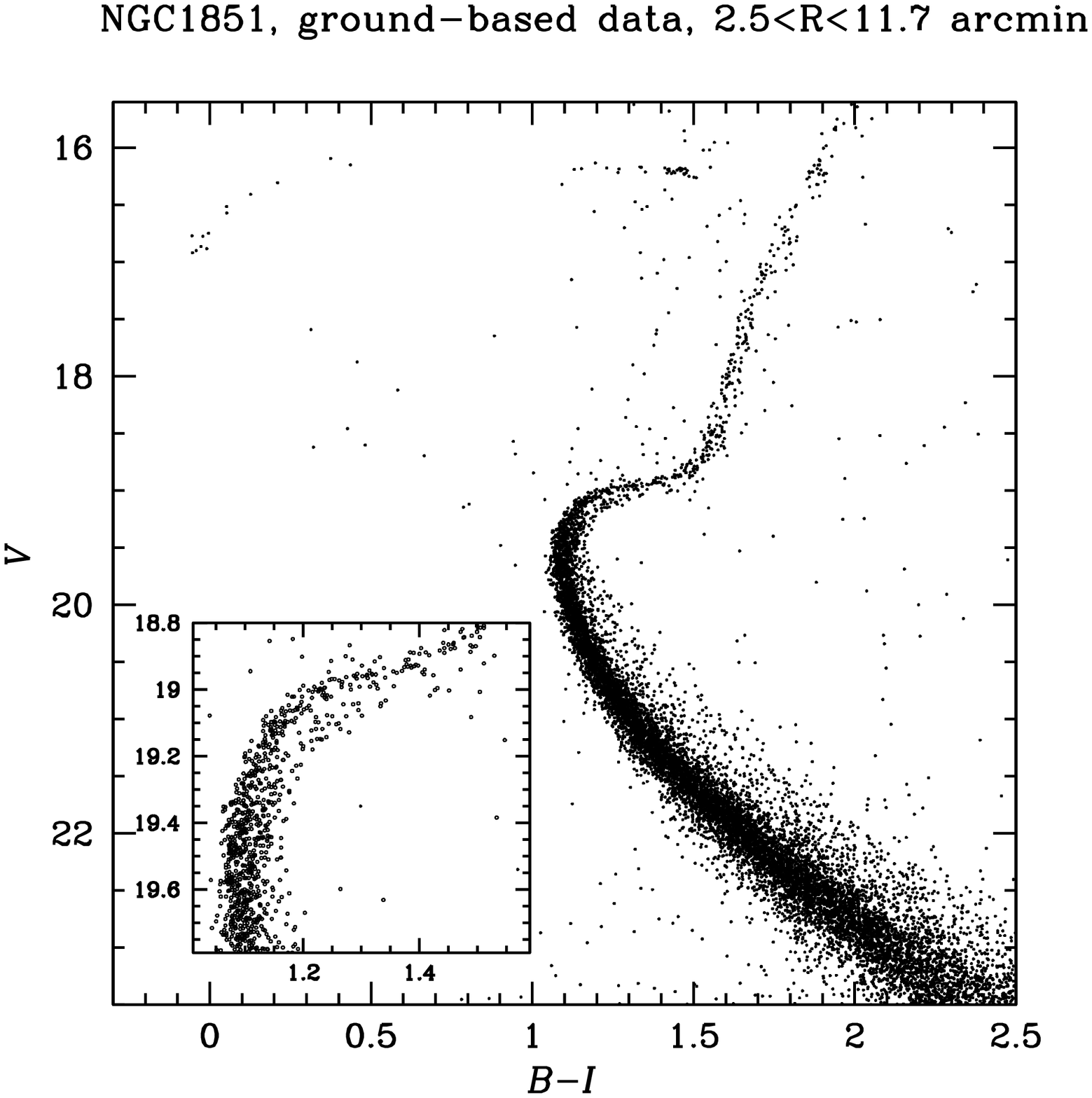}
      \caption{$V$ vs.\  $V-I$ ($left$)  and $V$ vs.\  $B-I$ ($right$)
        CMD of  NGC~1851 from ground-based data.  The inset show  a zoom around
        the SGB  region. Only stars with radial  distance greater than
        2.5 arcmin are plotted.  }
         \label{GB}
   \end{figure*}

\begin{figure*}[ht!]
   \centering
   \includegraphics[width=\textwidth]{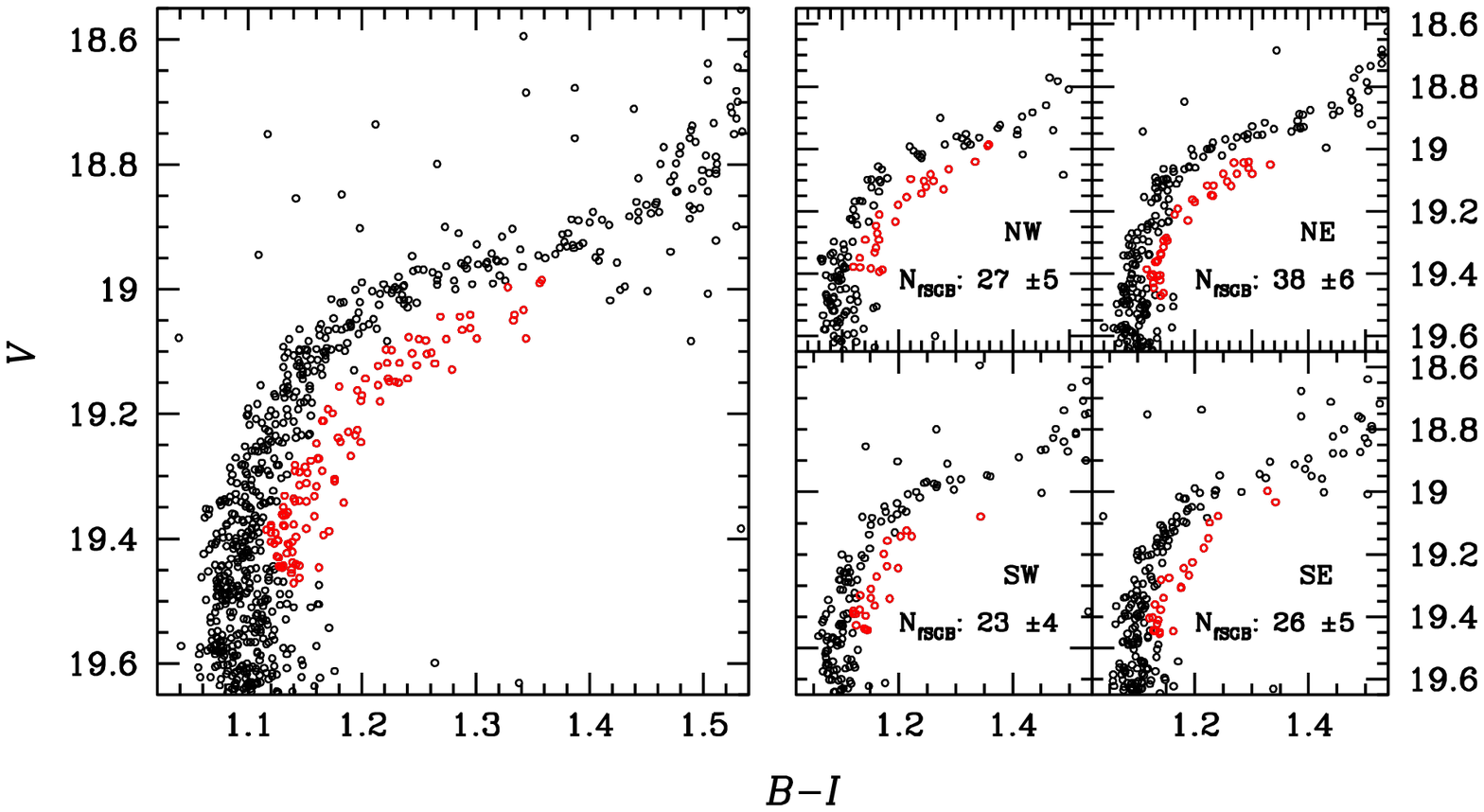}
      \caption{  ${\it V}$ vs.\  ${\it B-I}$ CMD of  NGC~1851
	from ground-based data for stars with radial distances from the cluster center larger than 2.5 arcmin ($left$) and CMDs 
	the
	four quadrants $right$. 
	}
         \label{manuela}
   \end{figure*}

The CMDs from ground-based    photometry are shown  in  Fig.~\ref{GB},
where we  have  rejected all  the stars   within 2.5 arcmin  from  the
cluster center.  The split of  the SGB is clearly  visible both in the
$(V-I)$ vs.\ $V$ and the $(B-I)$ vs.\ $V$ CMD, and the split region
is highlighted in
the inset where we show a zoom of the SGB region.
Note that the fact that we see the SGB split beyond $\sim$ 2.5 arcminutes
is already in disagreement with the claims by Zoccali et al.\ (2009)
who analysed the  S-W quadrant and found  no evidence for the presence
of fSGB stars at radial distances larger than  $\sim$ 2.4 arcmin.  The
left panel of  Fig.~\ref{manuela} 
shows our {${\it  V}$} vs ({${\it B-I} $})  CMD from ground based data
for stars with radial distance from the cluster center larger than 2.5
arcmin.  We marked in  red the stars  that more  likely belong to  the
fSGB.  In the four right panels we plot the same  CMD for stars in the
four quadrants.   The quoted numbers  are the number  of selected fSGB
stars with Poisson errors.  Because  of the uncertainties  we conclude
that  there is not significant  difference in the  distribution in the
CMD of fSGB stars in the four quadrants.  We detected 23 probable fSGB
stars in the S-W field covered by Zoccali et al. \ (2009) most of them
belonging to the  lower  part of the  fSGB.   Possibly, the difference
between our  result and the one  obtained by  Zoccali  et al  \ (2009)
comes from the small number of fSGB stars with radial distance greater
than $\sim$2.5  arcmin in the SW quadrant.   It must be noted that the
CMD of Zoccali  et al. \  (2009) have likely larger photometric errors
than in our case, because their photometry comes from a short (2s) and
a long (15s) FORS2-VLT image in ${\it V}$ and  ${\it I}$ bands.  Their
larger photometric errors  make  difficult the identification  of fSGB
stars especially in the lower part of the  SGB where the faint and the
bright SGB have a smaller color difference.

\begin{figure*}[ht!]
   \centering
   \includegraphics[width=\textwidth]{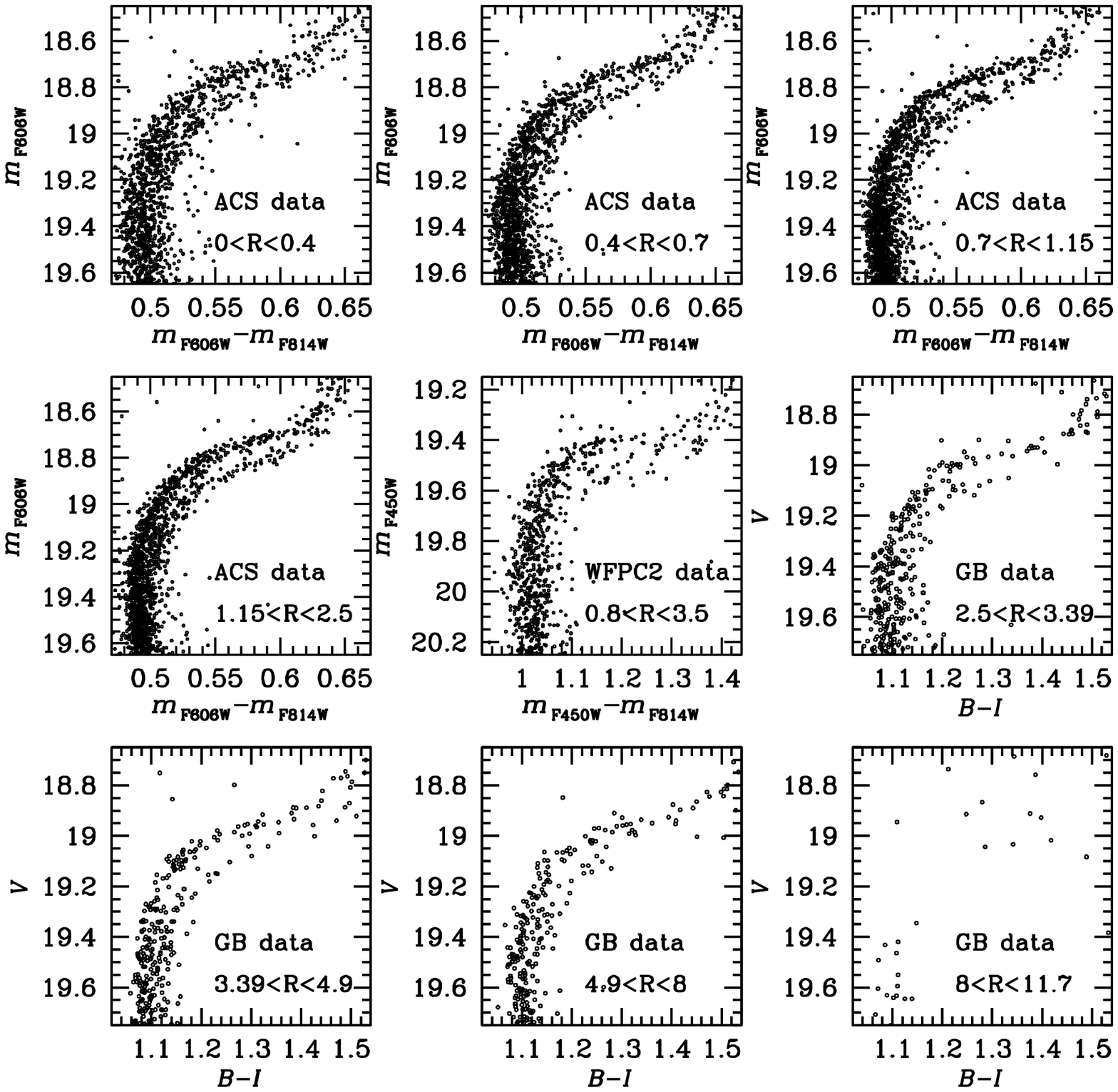}
      \caption{ CMDs of the region  around the SGB of NGC~1851 at different
        radial distances, from the center out to the tidal radius, and from the
        different data bases. The double SGB is visible out to where there are enough stars to see the SGB.
}
         \label{CMDs}
   \end{figure*}

The CMD in the right panel
of Fig.~\ref{GB} suggests that the RGB of NGC~1851
has some spread in the  $B-I$ color.  This spread could be related
to the presence of the two RGB branches observed by Calamida et al. (2007)
in the  Str\"omgren  ($m_{1}$,  $u-y$) CMD which is possibly associated
with the presence of two groups of stars with different CN abundances
(Yong \& Grundahl \ 2008).
%
To investigate how the  upper and lower  SGB populations may vary with
radius,  we divided the cluster  into   seven concentric annuli.   The
inner three are covered by the ACS data set out to 2.5 arcmin, and the
outer four are covered by the ground-based data set, going out to 12.5
arcmin.  In  Fig.~\ref{CMDs} we show the  CMD of the region around the
SGB for stars in each annulus.  Both the fainter and the brighter SGBs
are clearly visible    for   radial distances  smaller  than   $\sim$8
arcmin. Beyond this radius, the number of stars is simply too small to
distinguish  any feature of the SGB.   The WFPC2 observations allow an
additional check  on the  distribution of  stars along the  SGB in the
interval  between   0.8 and 3.5  arcmin  (see   the  central panel  in
Fig.~\ref{CMDs}).

\section{Radial distribution of the population ratio }

\label{RaD}
\subsection{The SGB subpopulations}
%

\begin{figure*}[ht!]
   \centering
   \includegraphics[width=\textwidth]{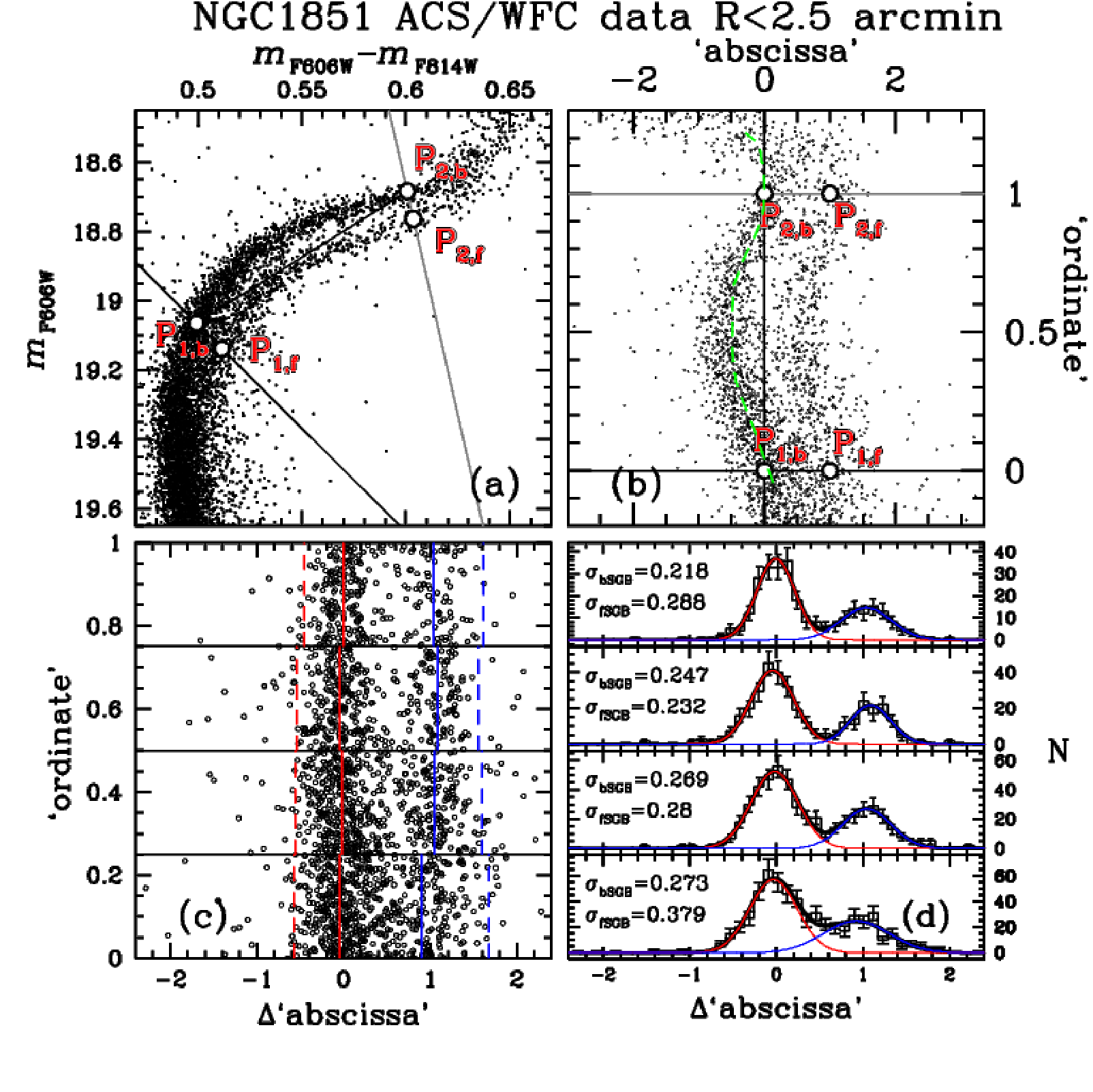}
      \caption{ This  figure  illustrates   the  procedure  adopted  to
        measure the fraction of stars belonging to the bSGB and fSGB
        in  NGC~1851.   Panel  (a)  shows  a  zoom  of  the  ACS/WFC CMD.
        The two  lines delimit the portion of the CMD where the split is
        most evident. Only stars
        from this region are used to measure the population ratio.  In
        Panel (b) we have transformed the reference frame of Panel (a).
        The green dashed line is  the fiducial of the region around the
        bSGB.  In Panel  (c) we plotted stars between  the two lines
        but after the subtraction of  the   fiducial line `abscissa'
	The  four right bottom panels  show the
        $\Delta$`abscissa'    distribution     for    stars    in    four
        $\Delta$`ordinate'   bins.  The   solid  lines   represent  a
        bigaussian  fit. For  each bin,  the dispersions  of  the best
        fitting Gaussian are indicated.
      }
         \label{PopRatioACS}
   \end{figure*}

To  determine the  fraction of   fSGB  and bSGB  stars,  we adopted  a
procedure  similar   to   that  used  by  Milone    at  al.\  (2009).
Fig.~\ref{PopRatioACS}  illustrates  this  four-step procedure for the
ACS/WFC sample.

We  selected by hand two points  on the fSGB ($P_{1,f}$,$P_{2,f}$) and
two points on the bSGB ($P_{1,b}$,$P_{2,b}$) with the aim of
delimiting  the SGB region   where the split   is most evident.  These
points define the two lines in panel (a), and  only stars contained in
the region between these lines were used in the following analysis.

In panel (b) we have  transformed the CMD linearly (the transformation
equation  is  given in  Appendix) into  a  reference frame  where: the
origin corresponds  to $P_{1,b}$; $P_{1,f}$ is  mapped into (1,0), and
the coordinates    of  $P_{2,b}$ and $P_{2,f}$  are    (0,1) and (1,1)
respectively.  For  convenience,  in the   following,  we indicate  as
`abscissa' and `ordinate'  the abscissa   and  the  ordinate of   this
reference frame. The dashed green line is the fiducial of the bSGB. We
drew it by  marking several points  on the  bSGB, and interpolating  a
line through them by means of a spline fit.  The  black lines of panel
(a) correspond to the loci with `abscissa' of  zero and one and to the
loci with `ordinate' of zero and one.

In panel (c) we have calculated the  difference between the `abscissa'
of each   star  and the  `abscissa' of  the    fiducial line ($\Delta$
`abscissa').

The  histograms  in   panel (d)  are   the distributions  in  $\Delta$
`abscissa'  for  stars  in four  $\Delta$ `ordinate'  intervals. These
distributions   have  been  modeled  as  the   sum  of  two  partially
overlapping  Gaussian functions.  To  reduce the influence of outliers
(such as stars with poor photometry,  field stars and binaries) we did
a  preliminary fit  of  the Gaussian  distribution using all available
stars.  Then we rejected all the stars more than two $\sigma_b$ to the
left of the bSGB and more than two $\sigma_f$ to the right of the fSGB
and repeated the fit  (the $\sigma$'s are   those of the  best fitting
Gaussian  in each $\Delta$`ordinate'  bin fitted to  the fSGB and bSGB
respectively).
In panel (c) the  continuous vertical lines indicate the  centers of  the best-fitting
Gaussians in  each $\Delta$`ordinate' interval.
The red dashed  line is located two $\sigma_{b}$  on the left  side of
the bSGB, and the blue dashed line runs two $\sigma_{f}$ on the right side of
the fSGB.
%

   \begin{figure*}[ht!]
   \centering
   \includegraphics[width=\textwidth]{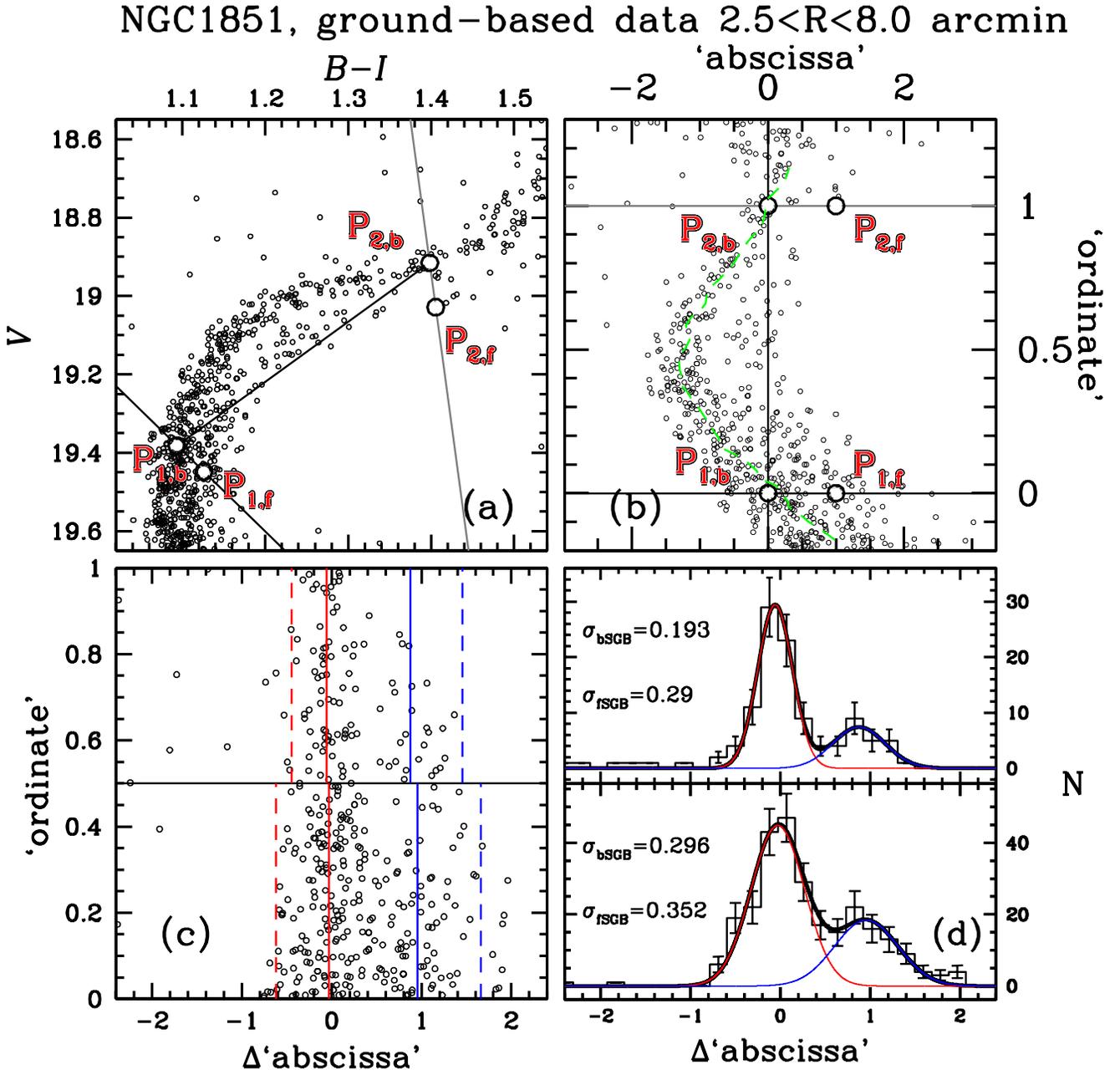}
      \caption{As in Fig.~\ref{PopRatioACS} for stars from the ground-based sample and with 2.5$<$R$<$8.0
        arcmin. }

         \label{PopRatioWFI}
   \end{figure*}
   \begin{figure*}[ht!]
   \centering
   \includegraphics[width=\textwidth]{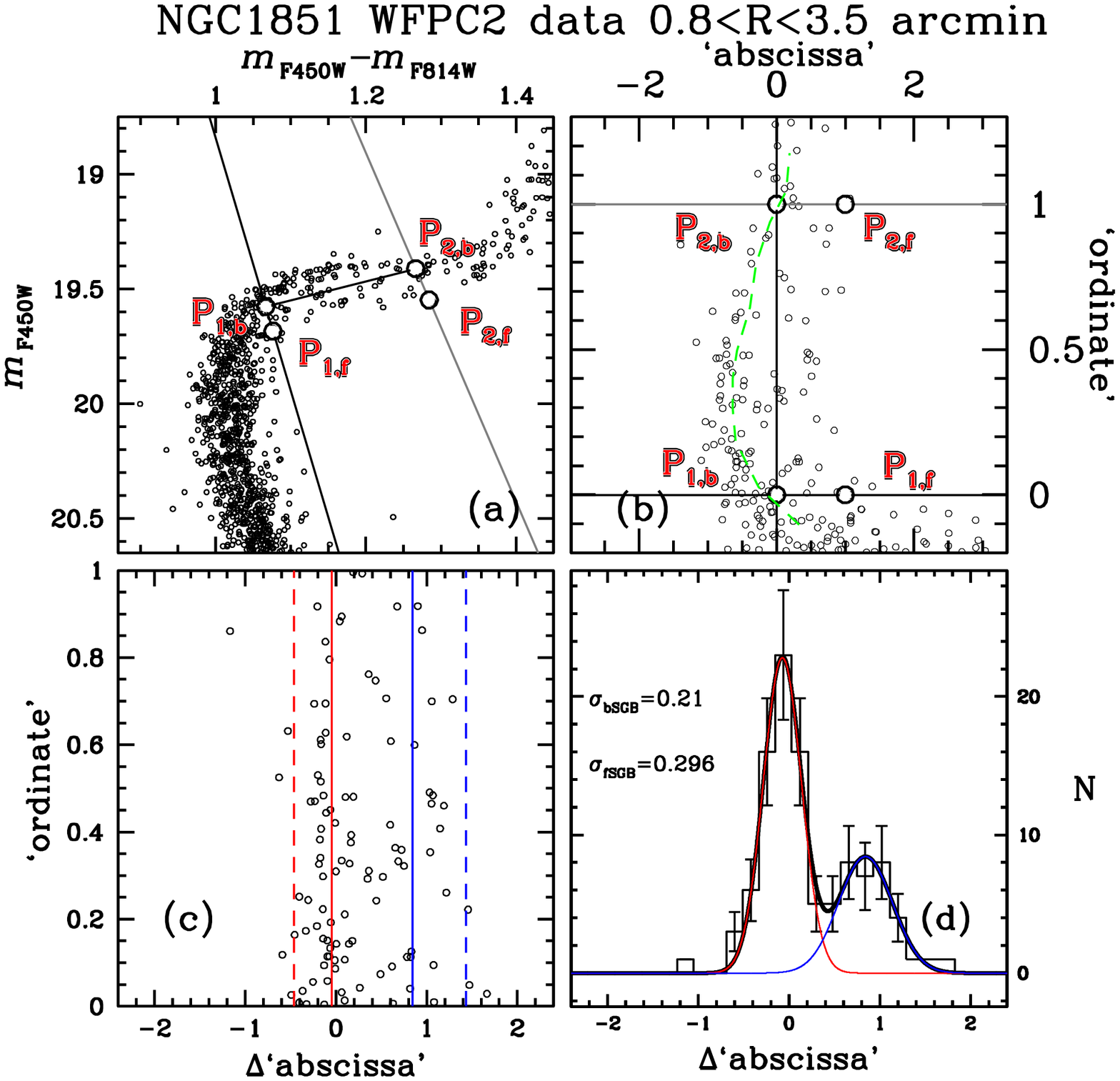}
      \caption{As in Fig.~\ref{PopRatioACS} for stars with 0.8$<$R$<$3.5
        arcmin. In this case we used the WFPC2
        photometry.  }
         \label{PopRatioWFPC2}
   \end{figure*}

We repeated the procedure for  the ground-based sample, as illustrated
in Figure~\ref{PopRatioWFI}.  In this case we  have reduced the number
of bins  in `magnitude' to just two  intervals, because of the smaller
number of stars.   The  same procedure has  also  been applied  to the
WFPC2 data.  In this  case,  we use    the entire  data set,   without
splitting it into bins, as illustrated in Fig.~\ref{PopRatioWFPC2}.

It is important to notice that each of the points
$P_{1,b}$,
$P_{1,f}$,
$P_{2,b}$, and
$P_{2,f}$---which we have arbitrarily defined
with the sole purpose of
isolating a group of stars representative of each of
the two SGBs---corresponds to a
different mass (
$\mathcal{M}_{P1b}$,
$\mathcal{M}_{P1f}$,
$\mathcal{M}_{P2b}$, and
$\mathcal{M}_{P2f}$).

To obtain a more accurate measure of the fraction  of stars in each of
the  two  populations (hereafter: $f_{\rm bSGB}$,  $f_{\rm fSGB}$)  we have to
compensate for the  fact that the two stellar  groups  that define the
two     SGBs      cover        two  different     mass       intervals
($\mathcal{M}_{P2f}-\mathcal{M}_{P1f}
\neq \mathcal{M}_{P2b}-\mathcal{M}_{P1b}$), due to the
different  evolutionary  lifetimes.   Consequently, the  correction we
have to apply will  be somewhat dependent on   the choice of the  mass
function.

To this end, we can calculate the fraction of stars in each branch as:
\begin{center}
$f_{\rm bSGB}  =  \frac { \frac {A_{b}} {N_{b}/N_{f}}}  {A_{f} +  \frac
    {A_{b}} {N_{b}/N_{f}}} $ \\
$f_{\rm fSGB}  =  \frac {A_{f}} {A_{f} +  \frac {A_{b}} {N_{b}/N_{f}}} $
\end{center}
where $A_{b}$ and $A_{f}$ are the  areas of the Gaussians that best fit
the bSGB and  the fSGB, and $N_{f(b)}=\int_{P_{1,f(b)}}^{P_{2,F(b)}}
\phi(\mathcal{M})d\mathcal{M}$.

As   for the  dependence  on  the  adopted mass function,   we ran the
following test. Because of the effect  of mass segregation, we assumed
a  heavy-mass-dominated  mass   function   for the  central    regions
($\alpha=-1.0$), and a steep ($\alpha=3.35$)   mass function for   the
external regions. Even with  these extreme assumptions, we found  that
the mass function effect can change  the relative fSGB/bSGB population
ratio by a negligible 4\%,  in the small  mass interval covered by our
SGB  stellar groups. Therefore, for simplicity,  we adopted a Salpeter
(1955) IMF for $\phi(\mathcal{M})$.

Since   the   inferred  population  ratio     depends on  the  assumed
evolutionary   lifetimes,    our   result   is    necessarily somewhat
model-dependent.  In fact, the values could even be scenario dependent
as well.  Cassisi et al.\ (2008) propose that the two SGBs in NGC~1851
could in principle be explained by one of the following scenarios:
\begin{itemize}
\item A) the bSGB and the fSGB are both populated by  two normal $\alpha$-enhanced
stellar populations with ages of 10 and 11 Gyrs, respectively;
\item B) the fSGB belongs to an 11 Gyr old, normal $\alpha$-enhanced stellar
population, while the bSGB belongs to a population 2 Gyr younger with high
C+N+O abundances;
\item C) the bSGB and the fSGB both correspond to a 10 Gyr old, normal $\alpha$-enhanced stellar
population, but the fSGB population has
C+N+O abundances a factor of two higher than the bSGB population.
\end{itemize}

Salaris et al.\  (2008) found  that scenario C  does the  best job  of
reproducing both the spectroscopic and photometric observations, while
scenario B  seems the  least consistent with   the  data.  Regardless,
these three possibilities illustrate that inferring the {\it actual\/}
frequency ratio   of     the populations depends    upon  the  assumed
astrophysical explanation of the difference  between them, through the
different   implications for the  evolutionary  lifetimes  between our
empirically chosen fiducial evolutionary states.

To better characterize the radial distribution of SGB stars we divided
both the ACS and ground-based data into concentric annuli as indicated
in Fig.~\ref{fields}.  The radial intervals  have been chosen so  that
the each of the four ACS  bins have the  same number of stars, and the
three ground-based bins   also have the same   number of stars   among
themselves.  In each of these  annuli,  we calculated the fraction  of
bSGB  and     fSGB   stars  using the      procedure  described above.
Table~\ref{tabfra} summarizes the results. The  first two columns list
the  radial interval  and  the average radial  distance  for SGB stars
within  this radial interval ($R_{AVE}$).  The  origin of the data and
the  number of SGB stars in  each bin are  indicated  in the third and
forth columns,  respectively,  while the  last three  columns list the
ratio  of faint to  bright SGB  stars that  we obtain for  each of the
three scenarios of Cassisi et al. (2008).

\begin{table*}[ht!]
\caption{Percentage of fSGB and bSGB stars at several radial distances. }
\scriptsize {
\begin{tabular}{ccccccc}
\hline\hline $R_{MIN}$ - $R_{MAX}$ &  $R_{AVE}$ & $INSTR$ & $N_{SGB}$
& $N_{fSGB}/N_{bSGB}$ (A)& $N_{fSGB}/N_{bSGB}$ (B)& $N_{fSGB}/N_{bSGB}$
(C) \\ \hline
0.0-2.5 & 0.80 & ACS   & 1746 &  0.60$\pm$0.03 & 0.69$\pm$0.03 & 0.53$\pm$0.03 \\
0.8-3.5 & 1.80 & WFPC2 &  105 &  0.56$\pm$0.09 & 0.52$\pm$0.09 & 0.54$\pm$0.09 \\
2.5-8.0 &      & G-B   &  303 &  0.49$\pm$0.07 & 0.53$\pm$0.07 & 0.51$\pm$0.07 \\
\hline \hline
0.0-0.4 &  0.28 & ACS  & 443  & 0.59$\pm$0.06 & 0.69$\pm$0.07 & 0.54$\pm$0.05 \\
0.4-0.7 &  0.57 & ACS  & 444  & 0.57$\pm$0.05 & 0.67$\pm$0.06 & 0.53$\pm$0.05 \\
0.7-1.2 &  0.89 & ACS  & 443  & 0.53$\pm$0.05 & 0.62$\pm$0.06 & 0.49$\pm$0.05 \\
1.2-2.5 &  1.54 & ACS  & 443  & 0.50$\pm$0.05 & 0.58$\pm$0.06 & 0.46$\pm$0.05 \\
0.8 3.5 &  1.80 & WFPC2& 105  & 0.56$\pm$0.09 & 0.52$\pm$0.09 & 0.54$\pm$0.09 \\
2.5 3.4 &  2.98 & G-B  & 100  & 0.41$\pm$0.08 & 0.44$\pm$0.08 & 0.40$\pm$0.08 \\
3.4 4.9 &  4.08 & G-B  & 101  & 0.47$\pm$0.10 & 0.55$\pm$0.11 & 0.49$\pm$0.10 \\
4.9 8.0 &  6.09 & G-B  & 100  & 0.50$\pm$0.11 & 0.54$\pm$0.12 & 0.52$\pm$0.11 \\

\hline
\end{tabular}
\label{tabfra}
}
\end{table*}

\begin{figure*}[ht!]
   \centering
   \includegraphics[width=11.5cm]{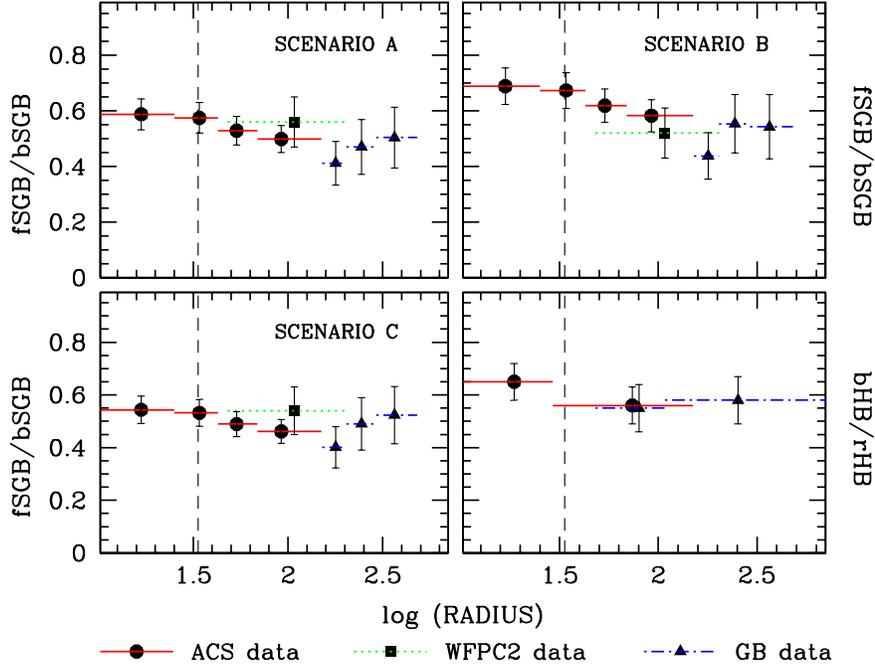}
      \caption{Fraction of fainter over  brighter SGB stars for the scenarios
	$A$, $B$, and $C$ and  fraction  of  blue   HB  over  red  HB.
	Circles, squares and triangles refer to the ACS/WFC, WFPC2 and WFI data
        sets.  The  dashed  vertical  lines  mark  the  core  and  the
        half-mass radius. }
         \label{RD}
   \end{figure*}

Since   the  values of  the  subpopulation  ratios  depend on the mass
interval corresponding   to  each SGB  segment,   the  results  differ
slightly depending on whether we assume that the SGB split corresponds
to scenario A, B, or C, as we use stellar  masses from three different
sets  of isochrones.  The  main result of  the  paper is summarized in
Fig.~\ref{RD}.   Although Scenarios A  and B show  a  hint of a slight
radial gradient, it is not statistically significant.  Scenario C does
not even show a  hint   of a  gradient.    Even when we maximize   our
statistics by putting the ACS data in one bin, the WFPC2 photometry in
a second bin, and all the ground-based data in a  third bin (see first
three rows   in Table~2), there  is  no significant variation   in the
fSGB:bSGB ratio with cluster radius.

When comparing the three different scenarios,  we note that the radial
trends are  slightly   different.  This has  no   reason to  be, since
different evolutionary times should not  change the relative trends of
the ratios.  The problem  comes from the  fact that we are using three
different data-sets, and we can not  avoid small transformation errors
from theoretical to observational plane.  ACS, WFPC2, and  ground-based
data  require slightly different   corrections because we  are working
with three different filter  systems, and we have  used the  models to
infer the relevant  mass  ranges.  These uncertainties generate  small
systematic errors in  our  estimate of the fSGB:bSGB  ratio.  However,
these  errors are  well within our   estimated error bars,  and do not
change the main result of a flat trend.

  We want to  emphasize that the fractions of  faint to bright SGB
stars  calculated  in this  paper   and  listed in Table~2  accurately
account for the mass interval covered by  each SGB segment.  Therefore
they differ from the  numbers quoted in  Milone et al.  \ (2008) where
we  have  simply determined  the  number of    stars belonging to  two
segments of  fSGB  and bSGB  in a  given  ${\it m}_{\rm F606W}  - {\it
m}_{\rm F814W}$ color interval.

%
\subsection{The HB subpopulation}
\begin{figure}[ht!]
   \centering
   \includegraphics[width=8.5cm]{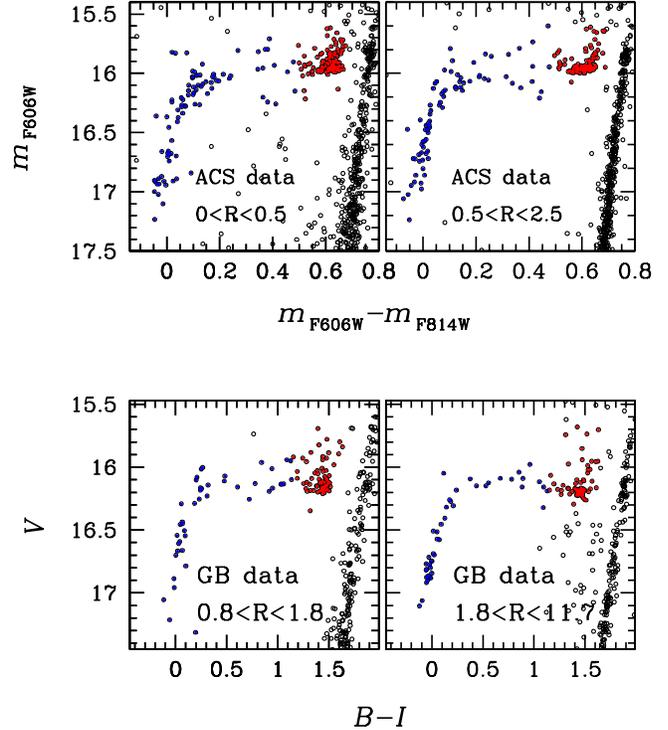}
      \caption{ Zoom of the CMD region around the HB at several radial
      distances from ACS (top) and WFI data  (bottom). We indicated in
      red  the probable red   HB stars and RR Lyrae   and in blue  the
      probable blue HB stars.
      }
         \label{HB}
   \end{figure}

As we mentioned above, NGC~1851 is  a prototypical bimodal HB cluster.
In Fig.~\ref{HB}  we show the HB  region for several annular bins both
from ACS/WFC (top panels) and  ground-based (bottom panels) data.  Red
HB stars are marked with red symbols, while blue HB stars and RR~Lyrae
are indicated in blue.

Table~3 lists  the ratio between  the blue HB stars  plus RR Lyrae and
red HB stars in three  different radial annuli.  In calculating  these
numbers we were careful to account  for the fact  that the lifetime of
star in the  blue HB is on average  11\% greater than the lifetime  in
the red  HB.  This estimate  has been  obtained  by comparing the core
He-burning  lifetime for the mean stellar  mass populating  the red HB
with that of  the mean stellar  mass populating the blue  HB in the HB
synthetic models  made by Salaris et al.  (2008) for simulating the HB
distribution in NGC1851.

The small number of intervals into which we have  divided the field of
view is a consequence  of  our need  for a statistically   significant
number   of stars   in each   subsample.  The  bottom right  panel  of
Fig.~\ref{RD} shows the  trend of  the (blue  HB +  RR~Lyrae):(red HB)
ratio as a function of the radial distance from the cluster center.

Note that for these   bright objects we   can now safely   include the
ground-based   photometry within 2.5 arcmin  from  the cluster center,
going as    close as 0.8 arcmin,    before crowding  conditions become
prohibitive even for these luminous stars.  The overlap regions of the
two data-sets,  between  0.8  and 2.5 arcmin,    provide  us with   an
important cross-check  on  the  solidity  and consistency of  the  two
independent sets of photometry.

Fig.~\ref{RD} shows that     there is no    statistically  significant
evidence of a radial  gradient in either   the fSGB:bSGB ratio  or the
bHB:rHB ratio.  Moreover, the results shown  in Fig. ~9 provide further
support to  the  suggestion of Milone et  al.\  (2008) and Salaris et
al.\ (2008) that the   blue  HB stars  are  the  progeny of the   fSGB
population, whilst the bulk of the red HB  component is related to the
bSGB  population.   More accurate   spectroscopical  measurements  are
mandatory in order to fully assess this scenario.

\begin{table}[ht!]
\scriptsize {
\begin{tabular}{cccccc}
\hline\hline $R_{MIN}$ - $R_{MAX}$ & $R_{AVE}$ & N & $INSTR$ & $(HB_b+RR Ly)/HB_r$\\
\hline
0.00 - 0.49 & 0.31  & 212  & ACS & 0.65$\pm$0.07 \\
0.49 - 2.51 & 1.23  & 212  & ACS & 0.56$\pm$0.07 \\
0.80 - 1.80 & 1.33  & 119  & G-B & 0.55$\pm$0.09 \\
1.80 - 11.7 & 4.22  & 118  & G-B & 0.58$\pm$0.09 \\
\hline
\end{tabular}
\label{tabfraHB}
}
\caption{Percentage of fSGB and bSGB stars at several radial distances. }
\end{table}

%
\section{Discussion}
\label{discussion}
%

D'Ercole et al.\ (2008) have shown that if a second generation
       of stars is formed by material
coming from or polluted by a first generation, then we would expect these
       stars to be born in the core of a stellar cluster, where
a cooling  flow collects the gas
ejected by  the earlier population.  As  the cluster evolves dynamically,
the  two populations mix and
 the ratio of second over first generation
tends  to a constant
value in the inner part of the cluster.  Until mixing is complete, the
radial profile of  this ratio is flat in the  inner part and decreases
in the outer cluster regions.
By studying the radial profiles of the different populations,
       we might hope to see evidence of these initial gradients before
       dynamical relaxation washes them out.

Before this paper and the work by Zoccali et al.\ (2009), $\omega$ Centauri  was  the only  cluster  where the
radial  distribution  of  different  stellar sub-populations  had  been
analyzed.  In  $\omega$ Centauri the stellar  population associated with
the blue, more  metal rich MS is more  centrally concentrated than the
red, more metal poor one, with  the relative ratio of blue over red MS
star counts being quite constant within the cluster core beyond
$\sim$ 12 arcmin and increasing by a factor of two from
8 arcmin  (Sollima et al. \ 2007) to about 5
arcmin from  the center.   Then the blue MS  to red MS  ratio remains
constant in the cluster core (Bellini et al. 2009).

In  this paper,  we have investigated  the radial  distribution of  the two
stellar  populations associated  with  the double  SGB  of NGC~1851.  By
coupling {\it HST} and  ground-based data we followed the distribution
of the two  populations from the center out to  the tidal radius, both
on  the SGB and  the HB.    
 Salaris et al.  \ (2008) claimed that  the ratio of the two SGBs
is consistent  with the idea   that the the   progeny of  the fSGB  si
distributed  from  the  blue  to   the  red HB,   including the  whole
instability strip,   while the bright one should    be confined to the
red.  According to  the scenario  of   Salaris et al.    \ (2008), the
fraction of fSGB that evolves into the red HB  corresponds only to the
5\% of the total number of stars.

At variance with Zoccali et al.  (2009), we have clearly detected both
the brighter and the  fainter  SGB at   all radial distances,  out  to
$\sim$8 arcmin.  At  larger radii, the number  of SGB stars  is simply
too small to  detect any substructure in  the SGB.  We have determined
the ratio of fainter to  brighter SGB stars and,  unlike the case with
$\omega$ Centauri,  found that---within    the error  bars---the   two
stellar  populations    have  the  same    radial  distribution.  This
conclusion does not depend on which of the three scenarios proposed by
Cassisi et al. (2008) we assume to explain the observed SGB dichotomy.
The ratio  of  the blue HB  stars  to the red HB   stars also shows no
significant trend with cluster radius.

Whatever  the origin  and  formation  process of  the two  stellar
generations in  NGC~1851, now the  two groups seem to  be well mixed
within the cluster.  Because of the short relaxation  time of NGC~1851
(log $t_{rh}$ $\sim$ 8.8) this result may be not totally unexpected, as
the  cluster stars must be  at  least  partially  mixed.  Unlike
NGC~1851,  $\omega$  Centauri  has  a very long relaxation time (log $t_{rh}$
$\sim$  10),  and it must have   retained  some   information  on   the  radial
distribution  of the gas  from which  its multiple  stellar generations formed.
However,  only  a  detailed dynamical  model,  following  the suggestion of
D'Ercole  et al.\ (2008) and Decressin et al. (2007),
will answer
       the question of whether the gradient seen in Omega Cen
       and the lack of a gradient seen in NGC~1851 imply different
       initial radial distributions for the various populations.

\begin{acknowledgements}
The authors wish to thank Francesca D'Antona for useful discussion.
We also thanks the anonymous referee for her/his comments and suggestions 
which helped to strenghen the results presented in this paper.
APM and GP acknowledge support by PRIN2007 and ASI under the
program ASI-INAF I/016/07/0.
J.A. acknowledges support from STScI Grants
                    GO-10775 and GO-11233.
\end{acknowledgements}
\bibliographystyle{aa}

{\bf Appendix}
Upon request from the referee, we give here 
the transformation equations used to pass from panel (a) to panel (b) in
Figs.~6, 7, and 8. 
For simplicity we indicate the color and the magnitude as $C$ and $M$ respectively,
and the `abscissa' and the `ordinate' as $abs$ and $ord$.

	\begin{equation}
	abs=(CR-c2)/c3
	\end{equation}
	\begin{equation}
	ord=yr/c1
	\end{equation}
	where:\\
	 \begin{equation}
	 CR=  (C-C_{P1,b})\, cos{\theta}+(M-M_{P1,b})\, sin{\theta}
	\end{equation}
	 \begin{equation}
	 MR= -(C-C_{P1,b})\, sin{\theta}+(M-M_{P1,b})\, cos{\theta}
	\end{equation}
	\begin{equation}
	\theta=atan\frac{MR_{P1,b}-MR_{P2,b}}{CR_{P1,b}-CR_{P2,b}}
	\end{equation}
		\begin{equation}
	 c1= \frac {MR_{P2,b}-MR_{P2,f}} {CR_{P2,b}-CR_{P2,f}}  CR
	+ \frac{CR_{P2,b}MR_{P2,f}-CR_{P2,f}MR_{P2,b}} {CR_{P2,b}-CR_{P2,f}}
		\end{equation}
		\begin{equation}
	 c2= \frac{CR_{P1,b}-CR_{P2,b}}{ord_{P1,b}-ord_{P2,b}}  ord + \frac{ord_{P1,b}CR_{P2,b}-ord_{P2,b}  CR_{P1,b}} {ord_{P1,b}-ord_{P2,b}}
		\end{equation}
\scriptsize{
	\begin{equation}
	 c3= \frac {CR_{P1,f}-CR_{P2,f}-c2_{P1,f}+c2_{P2,f}} {ord_{P1,f}-ord_{P2,f}}  ord + \frac {ord_{P1,f}(CR_{P2,f}-c2_{P2,f})-ord_{P2,f}(CR_{P1,f}-c2_{P1,f})}{ord_{P1,f}-ord_{P2,f}}
	\end{equation}
	}

\end{document}